\documentclass[a4paper,12pt]{article}
\usepackage[T1]{fontenc}
\usepackage{babel}
\usepackage{graphicx}
\usepackage{cite}

\begin{document}
\begin{center}
{\Large \bf \hfill August 28, 2000 \\
~\\
Does the KARMEN time anomaly originate from a beam-correlated 
background?
}  

\vspace{1mm}

\normalsize
F.\ Atchison,
M.\ Daum$^{\ast}$,
P.-R.\ Kettle,
C.\ Wigger
\end{center}
\noindent
PSI, Paul-Scherrer-Institut, CH-5232 Villigen-PSI, Switzerland.\\
$^{\ast}$E-mail address: Manfred.Daum@psi.ch  \\
Tel.: +41 56 310 36 68; Fax: +41 56 310 32 94.\\
\begin{center}
{\bf ABSTRACT}
\end{center}

\vspace{-3mm}

\noindent
The time anomaly of neutrino events observed in the KARMEN 
detector at the ISIS facility is discussed.
We show that beam-correlated neutron events are not completely
suppressed by the KARMEN lower energy cut and could cause the observed 
deviation of the measured time distribution from an exponential
curve with $\tau = 2.2\,\mu$s. \\
{\bf PACS: 13.20.Cz, 14.60.Pq, 14.80.-j}
\newpage
\noindent
The KARMEN collaboration investigates 
reactions induced by neutrinos produced in the pulsed neutron spallation source
of the ISIS facility at the Rutherford Appleton
Laboratory. In 1995, they reported\cite{karmen1} an anomaly 
in the time distribution of their events clustered around 3.6\,$\mu$s
after beam-on-target with visible energy between 
11 and 35\,MeV.
Further evidence of events exceeding 
the expected exponential distribution 
characterized by the muon lifetime has since been reported\cite{karmen2}.
The speculative explanation was 
that these events could be generated by 
hitherto unobserved neutral massive $X$-particles
originating from the ISIS target via the rare pion decay process,
\begin{equation}
\pi^+ \rightarrow \mu^+ X.
\end{equation}
From the kinematics of the two-body decay of a pion at rest,
and the known masses of the pion and the muon\cite{pdg94},
the mass of $X$ was calculated to be 33.9\,MeV
using the time-of-flight information and the mean flight-path 
of 17.5\,m from the ISIS target to the KARMEN detector.
Since the detected energy of the anomalous events is found to be 
much greater than the kinetic
energy of the $X$-particle, 
these events are postulated to originate from $X$-particles which decay 
with visible energy within the detector via the reaction
$X \rightarrow e^+ e^- \nu$.
The KARMEN collaboration estimated the branching fraction for the 
decay mode (1) as a function of the
$X$-lifetime, with values of the branching fraction ranging over many
decades down to $\sim$10$^{-16}$\cite{karmen1}.

Soon after the initial KARMEN publication\cite{karmen1},
various hypotheses were put
forward as to the nature of the 
$X$-particle\cite{barger,guzzo,cs,gninenko,govaerts},
and several searches undertaken to look for the rare 
decay\,(1)\cite{daum,bilger,bryman,bilger2}, all with negative results.
More recently, additional experiments\cite{nutev,daum99,nomad} have
set even more stringent limits for the branching fraction of the decay mode (1).

In this letter we present arguments for the existence 
of a beam-correlated neutron background in the 
KARMEN detector that could contribute to an
excess of events\footnote{The number of excess 
events reported in Ref.\,\cite{karmen1}
is $83 \pm 28$.}. Our basis is that
the KARMEN detector can respond to the 10.8\,MeV $\gamma$-energy 
from neutron capture by $^{14}\mbox{N}$. Thermal neutrons will
originate from moderation of fast neutrons ($T_n < 10\,$MeV)
in the liquid scintillator.
These fast neutrons are produced from medium energy interactions 
($T_n > 100\,$MeV) in the KARMEN iron shield.

The liquid scintillator (PPP) used in the KARMEN detector 
consists of
75\,\% (volume) paraffin oil, 25\,\% pseudocumene and 2\,g/l
PMP (1-phenyl-3-mesityl-2-pyrazoline)\cite{karmen3,guesten}.
The chemical formulae of the components are
(i) paraffin oil, taken as C$_{11}$H$_{24}$,
(ii) pseudocumene C$_9$H$_{12}$,
(iii) PMP C$_{18}$H$_{20}$N$_2$.
The nuclear densities 
in the liquid scintillator material (in units of nuclei/\AA$^3$)
are 0.071 for hydrogen, 0.036 for carbon and 
$9.1 \cdot 10^{-6}$ for nitrogen leading to the ratio
of atomic abundances of
\begin{equation}
n_H : n_C : n_N \approx 7800 : 4000 : 1.
\end{equation}
It is this nitrogen that can capture neutrons 
via the reaction
\begin{equation}
^{14}\mbox{N} (n, \gamma)^{15}\mbox{N} 
\end{equation}
with an integral $\gamma$-energy of 10.8\,MeV.
The thermal neutron cross-section for this reaction
is $\sigma \approx $75\,mb.
The lower energy cut of events in the KARMEN 
analysis is 11\,MeV\cite{karmen1} with an energy resolution of 
$\sigma_E/E = 11.5\,\%/\sqrt{E}$\cite{karmen3} ($E$ in MeV),
which is 3.5\,\%  or 380\,keV for $E = 11$\,MeV.
Thus, the 10.8\,MeV $\gamma$-events 
from reaction\,(3) will be accepted with an efficiency of 
about 30\,\%; we note that
in the energy distribution of the excess events\cite{oehler},
about 40\,\% of the anomalous events,
which are selected from an energy window between
11 and 35\,MeV, are clustered at about 11\,MeV. 
If such a beam-correlated neutron background is present,
the subtraction of purely cosmic background events 
is not sufficient.

The data fit published by the KARMEN collaboration gives a value
for the time constant $ \tau_{\mu} = (2.62 \pm 0.18)\,\mu$s\cite{karmen1}
which
differs by more than two standard deviations from the 
expected value of the muon lifetime ($2.19703 \pm 0.00004\,\mu$s).
We have refitted the KARMEN data of Ref.\cite{karmen1}
with an additional background of the form:
\begin{equation}
y(t) = A \cdot e^{-t/\tau_{\mu}} + B,
\end{equation}
and fixing $ \tau_{\mu}$ to the value of $2.2\,\mu$s.
The result of our fit is shown in Fig.\,1. 
This fit has a $\chi^2$ of 16.4 for 17 DOF and
the values of the two free parameters found are
$A = (271 \pm 16)$ events/$0.5\mu$s, $B = (7.2 \pm 3.0)$ events/$0.5\mu$s. 
The value for $B$,
greater than zero by 2.4 standard deviations,
can be interpreted as an indication for the presence of an
additional background with $137 \pm 57$ events\footnote{A similar fit was made
to the combined data from KARMEN1 and KARMEN2\cite{oehler}.}.

From the thermal capture cross-sections of hydrogen (333\,mb), carbon (3.5\,mb)
and nitrogen (75\,mb for n,$\gamma$ and 1810\,mb for n,p)
and using the ratio of atomic abundances\,(2),
we deduce the capture cross-section of the scintillator to be 
about 220\,mb and the probability for reaction (3) to be
$P \approx 3 \cdot 10^{-5}$. This means 
that $ 1/P \approx 3 \cdot 10^4$ thermal neutrons would be 
needed to create photons of reaction\,(3) with 10.8\,MeV of 
deposited visible energy.
In order to pass the 11\,MeV cut in the KARMEN analysis, this number has
to be a factor of 3 larger, i.e.\ about $1 \cdot 10^5$.

Because of the neutron absorbers around each scintillator 
element (gadolinium and borated polyethylene)\cite{karmen1},
thermal neutrons cannot enter the KARMEN detector
and hence have to be produced by moderating fast
neutrons in the scintillator fluid itself.
Since the hydrogen density in 
the PPP is similar to that for water,
the moderation characteristics should be alike.
`Age theory'\cite{moeller1} predicts an average time for thermalization of 
a few microseconds.
The time variation of the background signal 
will be complicated by the 1/v-tailback of the capture cross-section.
Thus, the contributions to the background signal, will start
well before thermalization is complete.

The events of Fig.\,1 originate from 6560\,C, i.e.\ 
$4\cdot 10^{22}$ protons hitting the ISIS target\cite{karmen1}.
Each proton produces about 0.05 medium energy neutrons
per steradian at 90 degree production angle\cite{atchie}.  The KARMEN detector
covers $5 \cdot 10^{-3}$ of the solid angle. The
material between the target and the KARMEN detector provides about 
4600\,g$\cdot$cm$^{-2}$ of shielding\cite{karmen1,karmen3,atchie,boardman},
and the shielding length for 
medium energy (above $\sim$
100\,MeV) neutrons is in the range 120 to 
160\,g$\cdot$cm$^{-2}$\cite{shielding}.
This leads to the estimate 
that between $5 \cdot 10^7$ and $3 \cdot 10^3$ medium energy neutrons
enter the KARMEN detector.
Medium energy neutrons 
contribute 0.4\,\% of the total neutron flux escaping from an
iron shield\cite{atchie}, and 
98\,\% of all neutrons have an energy below 
10\,MeV.
With these numbers, we get the contributions to 
our probability estimate of a beam-correlated neutron background
to be detected in the KARMEN time window (10$\mu$s)
listed in Table\,1.

From this table, we calculate the suppression factor 
of beam-correlated neutrons to be between $1\cdot 10^{-24}$
and $2\cdot 10^{-20}$
for the two extremes of the shielding length. 
Thus, for a total number of protons ($4 \cdot 10^{22}$),
the corresponding expected number of neutron events
originating from a beam-correlated background is between
0.06 and 800 which brackets the number of background events
(137 $\pm$ 57).

For a more refined background estimate, one has to consider that
in a pulsed neutron spallation source, a beam-correlated
neutron background has a time structure originating from the 
duty cycle of the primary proton beam and the kinetic energy
of the neutrons on their way through the shielding. 
The neutron capture rate of gadolinium absorber dissolved in water
was measured as a function of time after injection 
of a fast neutron pulse\cite{moeller2} and corresponds
to a linear rise from zero to about 6\,$\mu$s followed by a 
constant rate at longer times\footnote{In fact, an exponential
fall with a time constant of $\sim 100\,\mu$s\cite{karmen3}
(the thermal neutron lifetime in the KARMEN scintillator) is to be expected.}.
The time variation of the neutron capture rate in the liquid
scintillator is expected to be similar.
This behavior has been approximated by the function 
\begin{equation}
y(t) = A \cdot e^{(-t/2.2\mu s)} + B \cdot t/ \tau_c
\end{equation}
for t<$\tau_c$ and 
\begin{equation}
y(t) = A \cdot e^{(-t/2.2\mu s)} + B 
\end{equation}
for t>$\tau_c$.
Here, $A$
is the initial rate of the exponential decay, $B$ is the beam-correlated 
background from reaction\,(3), and $\tau_c$ is the 
linear risetime of the neutron capture rate.
A fit to the data of Ref.\cite{karmen1}
results in a $\chi^2$ of 15.5 for 16 DOF. The fitted values of the 
free parameters were $A = (277 \pm 14$) events/$0.5\mu$s, 
$B = (7.3 \pm2.8$) events/$0.5\mu$s, and 
$\tau_c = (3.4 \pm 0.5)\,\mu$s. This fit, shown in Fig.\,2,
leads to
$120 \pm 42$ neutron events from a beam-correlated background.
Although the statistical significance is similar to the 
previous fit, it gives a possible explanation for an enhancement
in the spectrum due to the neutron capture rate
in the scintillator reaching its maximum some 
3 to 4\,$\mu$s after beam on target.

After the upgrade of the KARMEN shielding (1996/97), any 
beam-correlated background from neutrons should be further suppressed
due to additional polyethylene, veto scintillation counters, 
and iron shielding. 
Thus, the corresponding `time anomaly' in the KARMEN 
data\cite{karmen1,oehler} before 1997,
should be less significant in the data after 1997, as was presented in 
Ref.\cite{eitel}.

In conclusion, it can at present not be ruled out
that the number of KARMEN excess events and their time distribution
according to Ref.\cite{karmen1}
are consistent 
with the hypothesis of originating from a beam-correlated neutron background 
involving neutron capture in nitrogen via reaction\,(3).
A more elaborate analysis of the time structure of such a
neutron background\cite{karmen2,oehler}
is beyond the scope of this letter, however, we would like to 
mention the existence of deep minima in the 
total cross section of iron, e.g.\ 
around 81\,keV\cite{rohr}.
At this energy, neutrons travel through iron
with a mean free path of several meters with a velocity of 4\,m/$\mu$s
and have a TOF of $\sim 2\,\mu$s from the surface of the ISIS biological
shielding to the KARMEN detector. These neutrons have a time structure 
similar to that of the proton beam on the ISIS target and so could lead
to a localized enhancement of the beam-correlated background event-rate.

\vspace{5mm}
The authors thank G.\ H.\ Eaton, R.\ Eichler, R. Frosch, 
and V. Markushin for valuable discussions and interest.
This work was supported by the Schweizerischer Nationalfonds.

\newpage
\noindent
\subsection*{Figure Captions}
\noindent
Figure\,1:\\
KARMEN data of Ref.\cite{karmen1} fitted with the function
y(t) = $A \cdot e^{-t/2.2\mu s} + B$. 
The smooth line represents the fitting function.
The dashed line is an exponential with $\tau_{\mu} = 2.2\,\mu$s,
and the dotted line represents an additional background. 
The values of the fitted parameters 
are $A = (271 \pm 16$)\,events/$0.5\mu$s, 
$B = (7.2 \pm 3.0$)\,events/$0.5\mu$s; the $\chi^2$ of the fit 
is 16.4 for 17 DOF.
~\\
~\\
Figure\,2:\\
KARMEN data of Ref.\cite{karmen1} fitted with the function
y(t) = $ A \cdot e^{-t/2.2\mu s} + B \cdot t/\tau_c$ for $t \leq \tau_c$
and y(t) = $ A \cdot e^{-t/2.2\mu s} + B $ for $t > \tau_c$.
The smooth line represents the fitting function.
The dashed line is an exponential with $\tau_{\mu} = 2.2\,\mu$s,
and the dotted line represents the time dependence of
our estimate of neutron capture in the KARMEN scintillator from reaction\,(3).
The values of the fitted parameters are 
$A = (277 \pm 14$)\,events/$\mu$s, $B = (7.3 \pm 2.8$)\,events/$\mu$s,
and $\tau_c = (3.4 \pm 0.5)\,\mu$s; the $\chi^2$ of the fit
is 15.5 for 16 DOF.

\newpage

\begin{figure}[htbp]

\vspace{-1cm}

{\bf \caption{~}}

\vspace{1cm}

\begin{center}

\includegraphics[width=86mm]{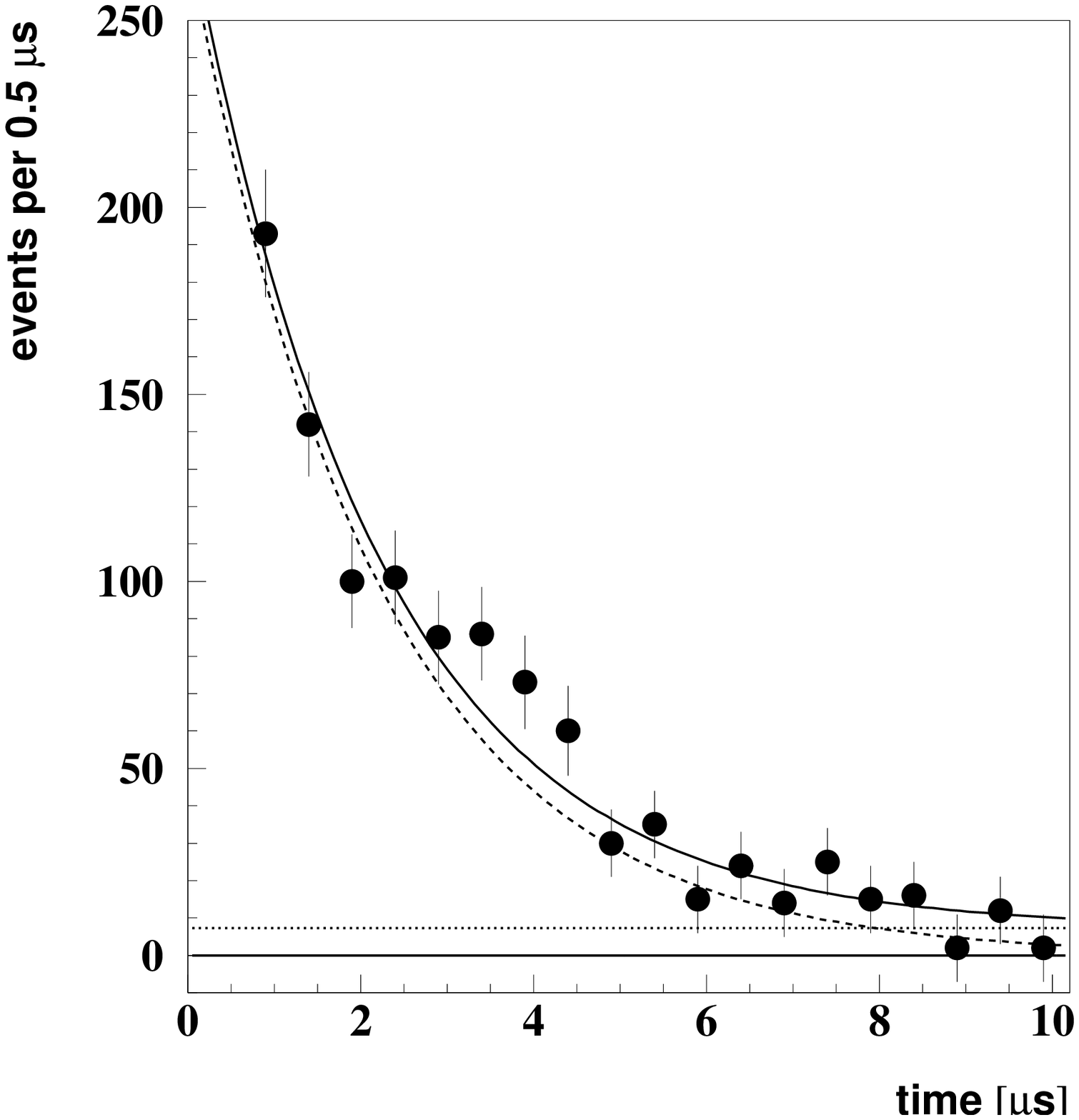}
\label{fi:setup}
\end{center}
\end{figure}

\begin{figure}[htbp]

\vspace{-1cm}

{\bf \caption{~}}

\vspace{1cm}

\begin{center}

\includegraphics[width=86mm]{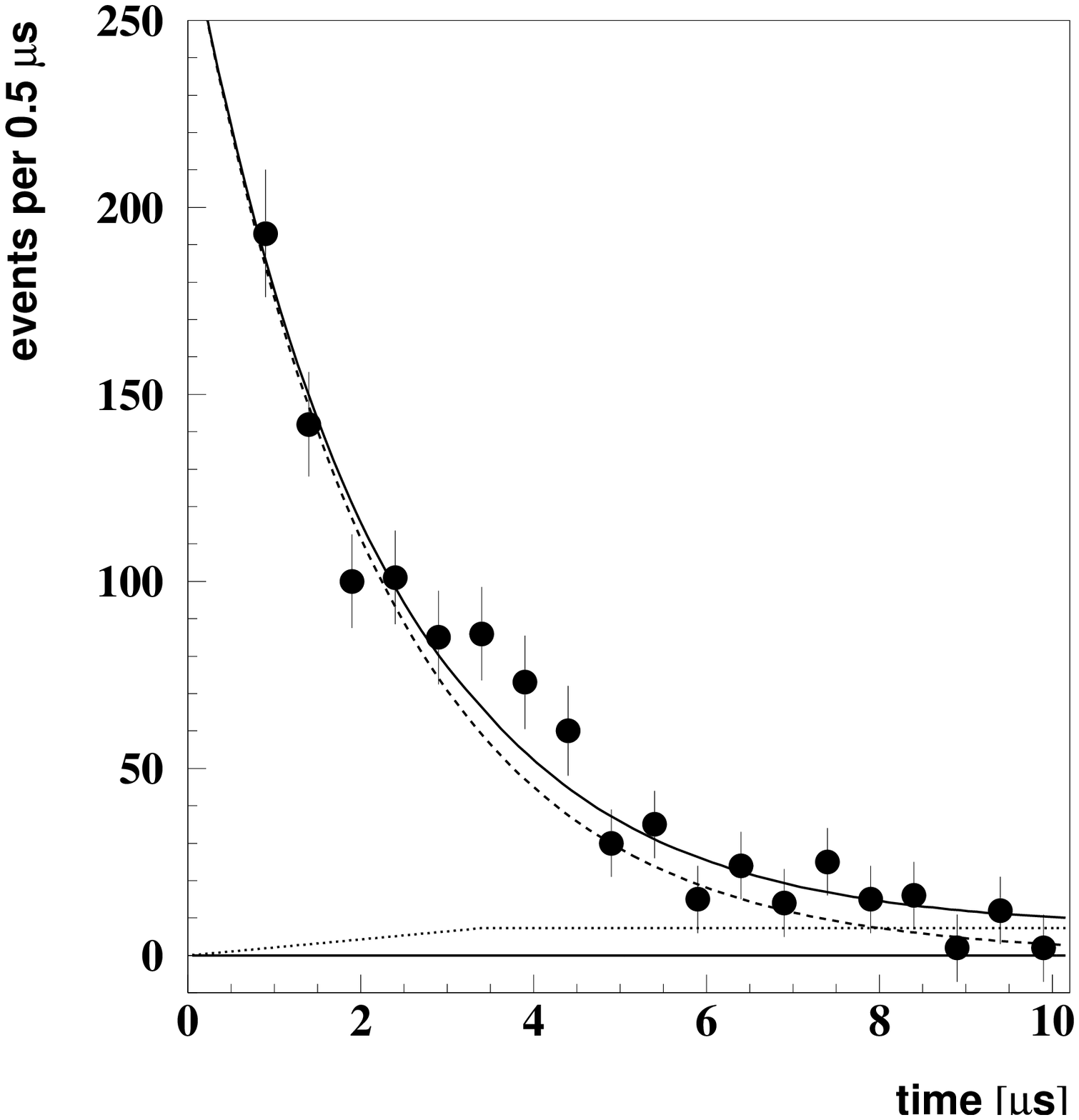}
\label{fi:setup}
\end{center}
\end{figure}

\newpage
\begin{table}
\caption{Factors used for estimating the beam-correlated neutron background}
\begin{tabular}{|l|c|}
\hline
Number of medium energy neutrons directed &\\
towards KARMEN & 0.05/sr/proton \\
\hline
Solid angle for the KARMEN detector & 0.0628\,sr \\
\hline
Shielding factors for 4600\,g/cm$^2$ and shielding lengths & \\
$\lambda$ of 120\,g/cm$^2 - 160$\,g/cm$^2$ & $2 \cdot 10^{-17} - 
3 \cdot 10^{-13}$ \\
\hline
Fraction of neutrons escaping the KARMEN steel shield & \\
in the medium energy region (0.4\,\%), & \\
in the fast neutron region (98\,\%); the number of fast & \\
neutrons per medium energy neutron is &  0.98/0.004 $\approx$ 250  \\
\hline
Shielding effect for fast neutrons of the borated & \\
polyethylene layer and the active outer shield & 0.25 \\
\hline
Fraction of neutrons captured by PPP & 0.78 \\
\hline
Probability of the capture channel 
$^{14}\mbox{N}(n,\gamma) ^{15}\mbox{N}$ &  $ 3 \cdot 10^{-5}$ \\
\hline
Detection efficiency for 10.8\,MeV 
(energy cut 11\,MeV) & 0.3 \\
\hline
Fraction of captures within the KARMEN time window & 0.05 \\
\hline
\end{tabular}
\end{table}

\end{document}